\documentclass[%
 aip,
 jmp,%
 amsmath,amssymb,
 reprint,%
]{revtex4-1}

\usepackage{graphicx}
\usepackage{dcolumn}
\usepackage{bm}

\usepackage{units}
\usepackage{color}

\begin{document}
\newcommand{\nuc}[2]{$^{#2}\rm #1$}

\newcommand{\bb}[1]{$\rm #1\nu \beta \beta$}
\newcommand{\bbm}[1]{$\rm #1\nu \beta^- \beta^-$}
\newcommand{\bbp}[1]{$\rm #1\nu \beta^+ \beta^+$}
\newcommand{\bbe}[1]{$\rm #1\nu \rm ECEC$}
\newcommand{\bbep}[1]{$\rm #1\nu \rm EC \beta^+$}

\newcommand{\rootcern}{\textsc{Root}}
\newcommand{\gerda}{\textsc{Gerda}}
\newcommand{\largeGERDA}{{LArGe}}
\newcommand{\PI}{\mbox{Phase\,I}}
\newcommand{\PIa}{\mbox{Phase\,Ia}}
\newcommand{\PIb}{\mbox{Phase\,Ib}}
\newcommand{\PIc}{\mbox{Phase\,Ic}}
\newcommand{\PII}{\mbox{Phase\,II}}

\newcommand{\geant}{\textsc{Geant4}}
\newcommand{\mage}{\myacs{MaGe}}
\newcommand{\decayzero}{\textsc{Decay0}}

\newcommand{\nPlus}{\mbox{n$^+$ electrode}}
\newcommand{\pPlus}{\mbox{p$^+$ electrode}}

\newcommand{\AOE}{$A/E$}

\newcommand{\order}[1]{\mbox{$\mathcal{O}$(#1)}}

\newcommand{\mul}[1]{\texttt{multiplicity==#1}}

\newcommand{\pic}[5]{
       \begin{figure}[ht]
       \begin{center}
       \includegraphics[width=#2\textwidth, keepaspectratio, #3]{#1}
       \caption{#5}
       \label{#4}
       \end{center}
       \end{figure}
}

\newcommand{\apic}[5]{
       \begin{figure}[H]
       \begin{center}
       \includegraphics[width=#2\textwidth, keepaspectratio, #3]{#1}
       \caption{#5}
       \label{#4}
       \end{center}
       \end{figure}
}

\newcommand{\sapic}[5]{
       \begin{figure}[P]
       \begin{center}
       \includegraphics[width=#2\textwidth, keepaspectratio, #3]{#1}
       \caption{#5}
       \label{#4}
       \end{center}
       \end{figure}
}

\newcommand{\picwrap}[9]{
       \begin{wrapfigure}{#5}{#6}
       \vspace{#7}
       \begin{center}
       \includegraphics[width=#2\textwidth, keepaspectratio, #3]{#1}
       \caption{#9}
       \label{#4}
       \end{center}
       \vspace{#8}
       \end{wrapfigure}
}

\newcommand{\baseT}[2]{\mbox{$#1\cdot10^{#2}$}}
\newcommand{\baseTsolo}[1]{$10^{#1}$}
\newcommand{\THL}{$T_{\nicefrac{1}{2}}$}

\newcommand{\UBI}{$\rm cts/(kg \cdot yr \cdot keV)$}

\newcommand{\Uflux}{$\rm m^{-2} s^{-1}$}
\newcommand{\Ucpd}{$\rm cts/(kg \cdot d)$}
\newcommand{\Uexpo}{$\rm kg \cdot d$}
\newcommand{\UexpoYear}{$\rm kg \cdot yr$}

\newcommand{\UMWE}{m.w.e.}

\newcommand{\Qbb}{$Q_{\beta\beta}$}

\newcommand{\validate}{\textcolor{blue}{\textit{(validate!!!)}}}

\newcommand{\improve}{\textcolor{blue}{\textit{(improve!!!)}}}

\newcommand{\missing}{\textcolor{red}{\textbf{...!!!...} }}

\newcommand{\quanta}{\textcolor{red}{\textit{(quantitativ?) }}}

\newcommand{\misscite}{\textcolor{red}{[citation!!!]}}

\newcommand{\missref}{\textcolor{red}{[reference!!!]}\ }

\newcommand{\PC}{$N_{\rm peak}$}
\newcommand{\BIC}{$N_{\rm BI}$}
\newcommand{\PAPR}{$R_{\rm p/>p}$}

\newcommand{\PCR}{$R_{\rm peak}$}


\newcommand{\gline}{$\gamma$-line}
\newcommand{\glines}{$\gamma$-lines}

\newcommand{\gray}{$\gamma$-ray}
\newcommand{\grays}{$\gamma$-rays}

\newcommand{\bray}{$\beta$-ray}
\newcommand{\brays}{$\beta$-rays}

\newcommand{\aray}{$\alpha$-ray}
\newcommand{\arays}{$\alpha$-rays}

\newcommand{\betas}{$\beta$'s}


\newcommand{\tab}{{Tab.~}}
\newcommand{\eq}{{Eq.~}}
\newcommand{\fig}{{Fig.~}}
\renewcommand{\sec}{{Sec.~}}
\newcommand{\chap}{{Chap.~}}

 \newcommand{\fn}{\iffalse \fi} 
 \newcommand{\tx}{\iffalse \fi} 
 \newcommand{\txe}{\iffalse \fi} 
 \newcommand{\sr}{\iffalse \fi} 

\preprint{AIP/123-QED}

\title{Search for the decay of natures rarest isotope $^{\mathbf{180m}}$Ta}

\author{B. Lehnert}%
\email{blehnert@physics.carleton.ca}
\affiliation{Institut f\"{u}r Kern- und Teilchenphysik, Technische Universit\"{a}t Dresden, Germany}
\affiliation{Present Address: Physics Department, Carleton University, Ottawa, Canada}

\author{M. Hult}
\email{mikael.hult@ec.europa.eu}
\affiliation{European Commission, JRC-Geel, Retieseweg 111, B-2440 Geel, Belgium}

\author{G. Lutter}
\email{guillaume.lutter@ec.europa.eu}
\affiliation{European Commission, JRC-Geel, Retieseweg 111, B-2440 Geel, Belgium}

\author{K. Zuber}
\email{zuber@physik.tu-dresden.de}
\affiliation{Institut f\"{u}r Kern- und Teilchenphysik, Technische Universit\"{a}t Dresden, Germany}
\date{\today}

\begin{abstract}
$^{\rm {180m}}$Ta is the rarest naturally occurring quasi-stable isotope and the longest lived metastable state which is known. Its possible decay via the $\beta^-$ or the electron capture channel has never been observed. This article presents a search for the decay of $^{\rm {180m}}$Ta with an ultra low background Sandwich HPGe gamma spectrometry setup in the HADES underground laboratory. No signal is observed and improved lower partial half-life limits are set with a Bayesian analysis to \unit[$5.8\cdot10^{16}$]{yr} for the $\beta^-$ channel and \unit[$2.0\cdot10^{17}$]{yr} for the electron capture channel (\unit[90]{\%} credibility). The total half-life of $^{\rm {180m}}$Ta is longer than \unit[$4.5\cdot10^{16}$]{yr}. This is more than a factor of two improvement compared to previous searches.

\end{abstract}

\pacs{Valid PACS appear here}
\maketitle

\section{Introduction}
The study of very long living nuclides is by itself a very interesting topic. 
Considering beta decays, this implies highly forbidden transitions. Examples are
\nuc{Cd}{113}, \nuc{In}{115} and \nuc{V}{50} which are 4-fold forbidden non-unique
decays. Even higher forbidden beta decays should exist for example in \nuc{Ca}{48}
and \nuc{Zr}{96} which can compete with double beta decay. \\

A very special candidate is \nuc{Ta}{180}. From all 300 stable nuclei only 9 are odd-odd nuclei and \nuc{Ta}{180} is the heaviest  one. Furthermore, it is the rarest quasi-stable isotope of the rare earth elements with an abundance
of only 0.01201(8)\% \cite{delaeter2005} and its production in the stellar nucleosynthesis has been debated for decades. The reason for this is that it is bypassed by the main production processes like r- and s-process (see for example \cite{mohr2007,hayakawa2010}).
Another unique feature of \nuc{Ta}{180} is it being the only quasi-stable isomer as the ground state decays with
a half-life of 8.1 hrs. The \nuc{Ta}{\rm 180m} is stabilized by its high spin of 9$^-$  due to the spin alignment
$\pi 9/2 [514] + \nu9/2 [624]$ while the ground state is anti-aligned  $\pi 7/2 [404] + \nu9/2 [624]$
resulting in a spin of $1^+$. Due to the high spin difference, a depopulation of the \nuc{Ta}{\rm 180m} can only
occur by photo-excitation into excited states which have a decay branch into the ground state.\\

\begin{figure}
\includegraphics[width=0.5\textwidth]{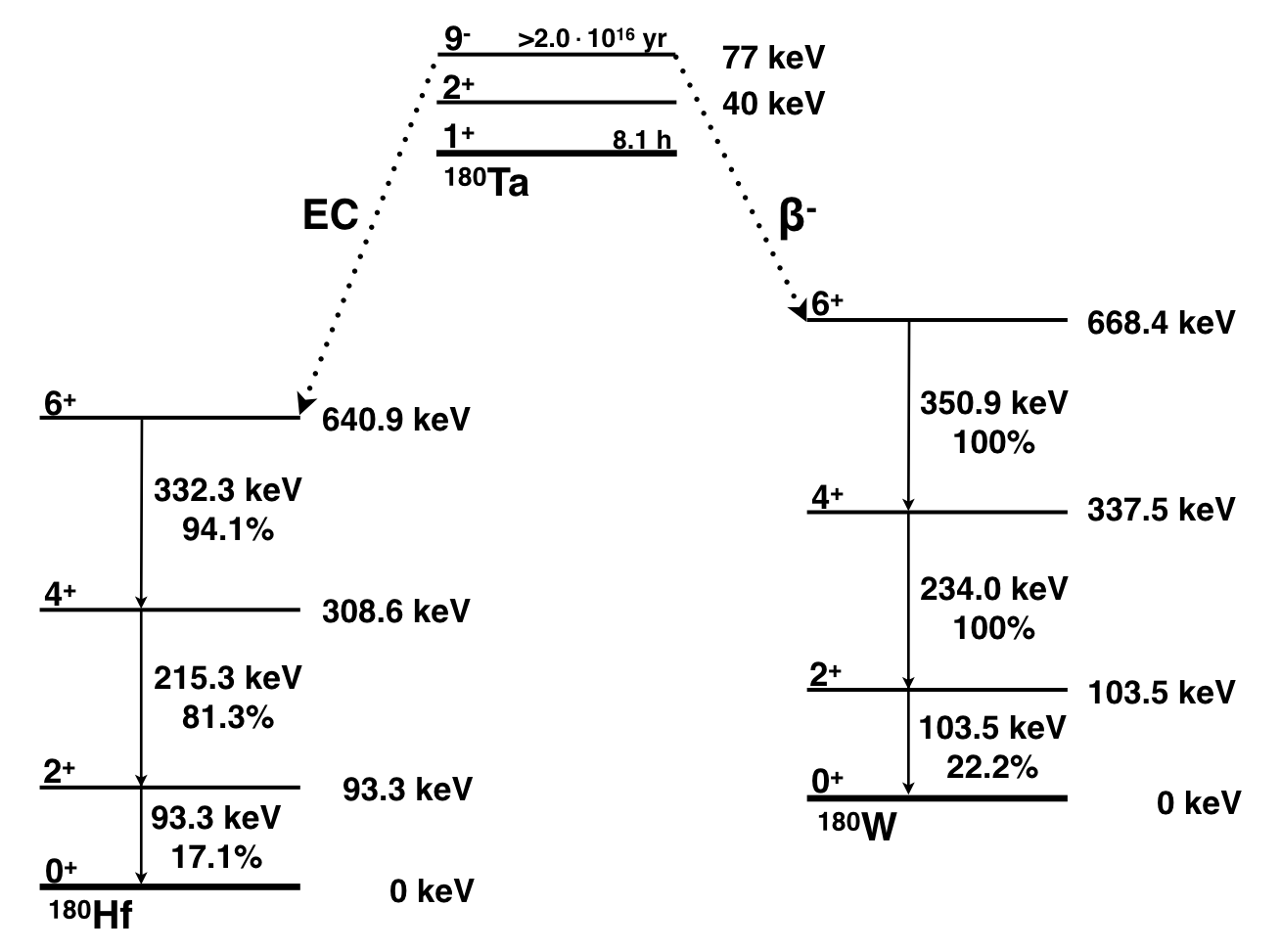}
\caption{\label{fig:decayScheme} Decay scheme of \nuc{Ta}{180} with data from \cite{NuclData}.}
\label{fig:decayscheme}
\end{figure}

The decay scheme of \nuc{Ta}{\rm 180m} is shown in Fig.~\ref{fig:decayScheme}. It branches with beta decay
to  \nuc{W}{180} and with electron capture (EC) to \nuc{Hf}{180}. The lowest spin difference is to the
excited 6$^+$ states which cascade down to the ground states. The transitions would be classified
as 3-fold forbidden non-unique like \nuc{Rb}{87} beta decay.\\

Several past measurements have been performed to search for the decay of \nuc{Ta}{\rm 180m} which are summarized in \tab \ref{tab:table1}. The best current half-life limits are \unit[\baseT{4.45}{16}]{yr} for the EC branch, \unit[\baseT{3.65}{16}]{yr} for the $\beta^-$ branch and \unit[\baseT{2.0}{16}]{yr} for the total half-life \cite{Ta108m2009}. \\

\begin{table*}
\caption{\label{tab:table1} Previous results on \nuc{Ta}{\rm 180m} half-lives. The various detection techniques are indicated with a simplified description. For measurements \cite{Norman81} and \cite{Cumming85} an enriched tantalum sample was used.}
\begin{ruledtabular}
\begin{tabular}{llcrrr}
Reference  &  Year  &  Technique  & \multicolumn{3}{c}{Lower half-life limit}\\
  &    &    & EC & $\beta^-$ & total\\
\hline
Eberhardt et al. \cite{Eberhardt55}      &  1955     & Mass Spec.                               &         ---                  & \unit[\baseT{9.9}{11}]{yr}   & ---\\
Eberhardt and Signer \cite{Eberhardt58}  &  1958     & Mass Spec.                               & \unit[\baseT{4.6}{9}]{yr}    & ---                          & \unit[\baseT{4.5}{9}]{yr}  \\
Bauminger and Cohen \cite{Bauminger58}   &  1958     & $\gamma$-spec.\ NaI                      & \unit[\baseT{2.3}{13}]{yr}   &  \unit[\baseT{1.7}{13}]{yr}  & \unit[\baseT{9.7}{12}]{yr} \\
Sakamoto \cite{Sakamoto67}               &  1967     & $\gamma$-spec.\ NaI                      & \unit[\baseT{1.5}{13}]{yr}\ $^a$   & ---                          & --- \\
Ardisson \cite{Ardisson77}               &  1977     & $\gamma$-spec.\ Ge(Li)                   & \unit[\baseT{2.1}{13}]{yr}   &   ---                 & --- \\
Norman  \cite{Norman81}                  &  1981     & $\gamma$-spec.\ Ge(Li) enr.\ Ta 	        & \unit[\baseT{5.6}{13}]{yr}   &  \unit[\baseT{5.6}{13}]{yr}  & \unit[\baseT{2.8}{13}]{yr} \\ 
Cumming and Alburger  \cite{Cumming85}   &  1985     & $\gamma$-spec.\ HPGe enr.\ Ta            & \unit[\baseT{3.0}{15}]{yr}   &  \unit[\baseT{1.9}{15}]{yr}  & \unit[\baseT{1.2}{15}]{yr} \\
Hult et al. \cite{Ta108m2006}            &  2006     & $\gamma$-spec.\ HPGe                     & \unit[\baseT{1.7}{16}]{yr}   &  \unit[\baseT{1.2}{16}]{yr}  & \unit[\baseT{7.2}{15}]{yr} \\                                                                                                                    
Hult et al. \cite{Ta108m2009}            &  2009     & $\gamma$-spec.\ Sandwich HPGe            & \unit[\baseT{4.45}{16}]{yr}  &  \unit[\baseT{3.65}{16}]{yr}  & \unit[\baseT{2.0}{16}]{yr}                                                                                                                     
\end{tabular}
\end{ruledtabular}
$^a$ Indication for a positive signal was given.
\end{table*}

The present study is in principle an extension of the previous measurement \cite{Ta108m2009} using the same sample and detectors but incorporates in addition the following features:

\begin{itemize}
\item 176 more days of measurement time.

\item Reduced intrinsic background due to further decay of cosmogenic radionuclides in the sample, detectors and shield.

\item Improved statistical analysis based on spectral fits and calculating the full Bayesian posterior probability.

\item Combining multiple \grays\ of the de-excitation cascade in a given decay mode.

\item Combining multiple data sets incorporating all the previous measurements performed in the underground laboratory HADES including the study from 2006 \cite{Ta108m2006} using another tantalum sample.

\end{itemize}


\section{Sample}

\begin{figure}
\includegraphics[width=0.4\textwidth]{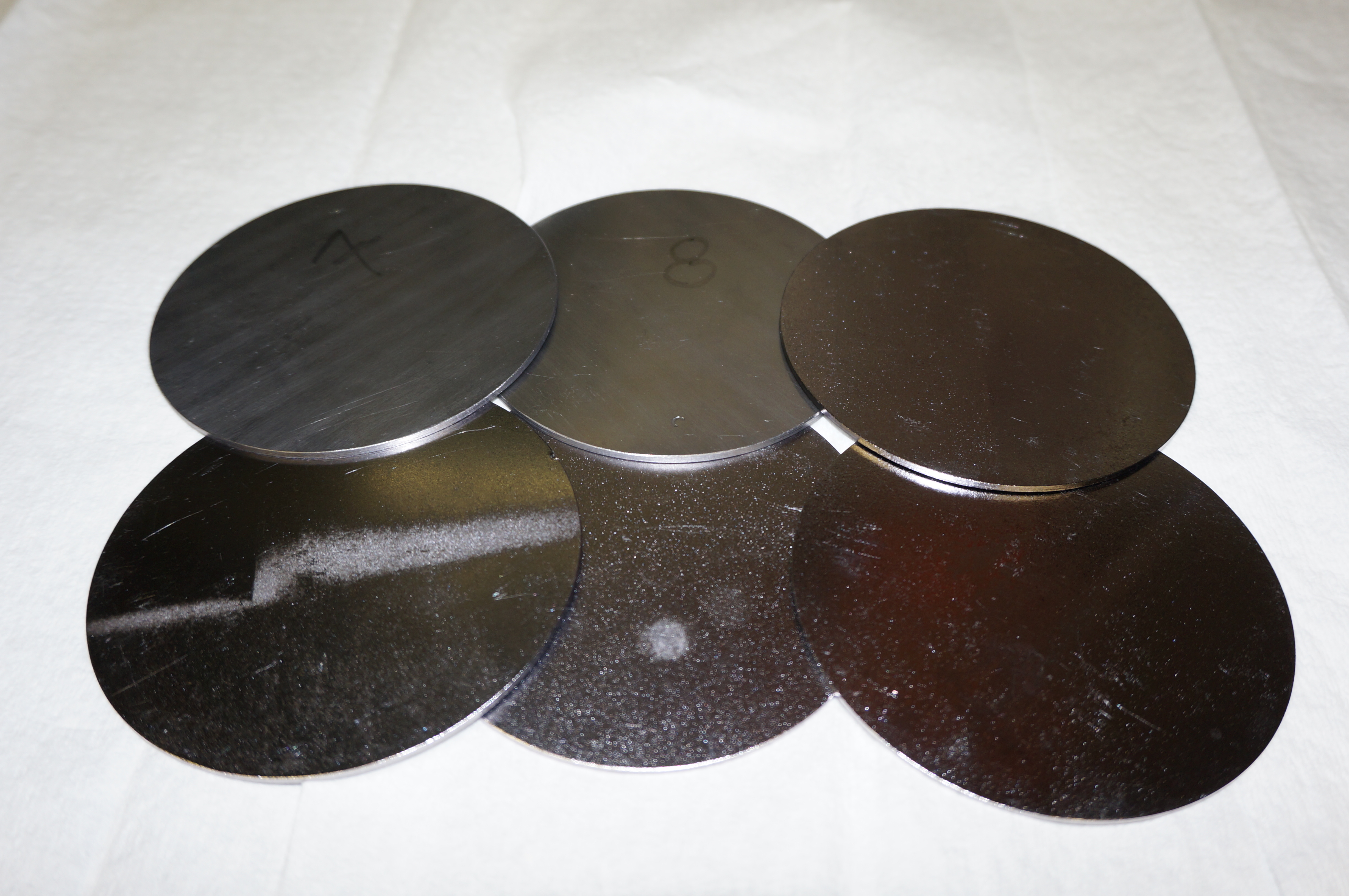}
\caption{\label{fig:sample} Picture of the tantalum sample.}
\end{figure}

The natural isotopic abundance of \nuc{Ta}{180} is very low and it is very cumbersome and therefore expensive to produce a sample enriched in \nuc{Ta}{\rm 180m}. Cumming and Alburger \cite{Cumming85} used an enriched sample containing \unit[8]{mg} of \nuc{Ta}{\rm 180m}, which is relatively small and which has the additional problem of not being perfectly radiopure. For this study, the same sample as in \cite{Ta108m2009} was used which is shown in \fig \ref{fig:sample}. This sample consists of 6 disks with \unit[100]{mm} diameter and \unit[2]{mm} thickness of high purity tantalum of natural isotopic composition. The disks have a mass of \unit[1500.33]{g} translating into a total \nuc{Ta}{\rm 180m} mass of \unit[180]{mg}. The disks have been underground in HADES for more than 6 years prior to these measurements, which guarantees that e.g. the activity of \nuc{Ta}{182} (\unit[$T_{1/2} = 114.4$]{d}) has decreased to insignificant levels and do not disturb these measurements. 

\section{Setup}

\begin{figure}
\includegraphics[scale=0.3]{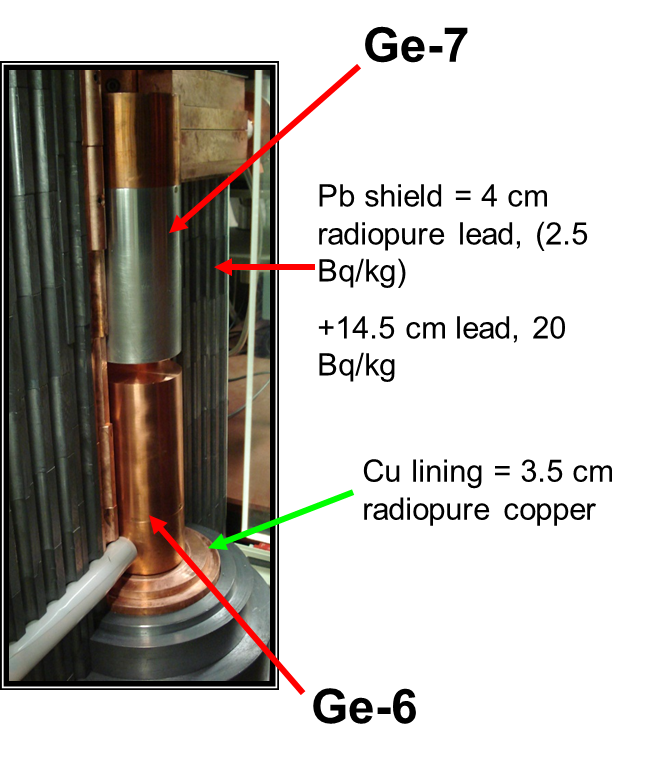}
\caption{\label{fig:setup} Picture of the Sandwich spectrometer.}
\end{figure}

The measurements took place in the laboratory HADES, which is located \unit[225]{m} underground at the premises of the Belgian nuclear centre SCK$\cdot$CEN in Mol, Belgium. In the underground location, the atmospheric $\mu$-flux is reduced by a factor 5000 compared to the surface. JRC-Geel operates a range of HPGe-detectors in HADES as part of its radionuclide metrology laboratory. The detectors are used in a wide range of applications stretching from characterization of reference materials and environmental radioactivity studies to measurements of decay data. The shape of the Ta-sample makes it highly suitable for measurement in the so-called Sandwich detector system \cite{Wieslander09}. It consists of two coaxial HPGe-detector, Ge-6 and Ge-7. Detector Ge-6 has a normal U-style arm, whilst detector Ge-7 has an arm that is rotated 180 deg so that its endcap is facing down. A picture of the arrangement is shown in \fig \ref{fig:setup}. Detector Ge-7 was moved downwards so that the distance between the two detector-endcaps was only \unit[1.3]{cm} for these measurements. Detector Ge-6 has a relative efficiency of \unit[80]{\%} and has a thick (\unit[0.5]{mm}) upper deadlayer, whilst Ge-7 has a relative efficiency of \unit[90]{\%} and a thin (\unit[0.3]{$\mu$m}) upper deadlayer.

An active muon shield made of two plastic scintillator (PS) is installed on the top of the lead shield. The coincidence signal from the two PS is used as a hardware gating signal. This signal and the signals of the two HPGe-detectors are sent to standard NIM modules (amplifiers and ADCs) connected to a multi-parameter system called DAQ2000\cite{DAQ2000}. The DAQ2000 is based on LabView\copyright~ and designed and manufactured by JRC-Geel. The events are time-stamped with 100 ns binning. All the events are stored in list-mode and the anti-coincidence is set by software during the offline analysis. In parallel, the HPGe data are also collected using the Canberra "Genie-2000" acquisition system without muon veto coincidence.

The full energy peak (FEP) efficiencies were calculated using the Monte Carlo code EGSnrc \cite{EGSnrc}. The two HPGe-detectors have been modeled by using physical dimensions provided by manufacturers and extracted from radiographs. The values of the position and the dead layer of the crystal were adjusted until a good agreement (better than \unit[5]{\%}) between the measured and calculated efficiencies was obtained. The Monte Carlo model of the two detectors was also validated through the several participations in proficiency testing schemes.
The two decay branches of \nuc{Ta}{\rm 180m} were simulated separately using the nuclear data from \cite{NuclData}. All gamma-rays and X-rays of the decay scheme were considered so that the calculated FEP efficiencies intrinsically included the coincidence summing effect. Angular correlations were neglected and the simulations assumed that the activity was homogeneously distributed in the sample.


\section{Analysis}

The analysis is performed as a spectral fit for the $\beta^-$ and EC decay mode independently. The fit is performed with a single half-life parameter on four individual datasets $d$ named "M1" for the 2006 publication\cite{Ta108m2006}, "M2" for the 2009 publication\cite{Ta108m2009} and "M3 Ge6" and "M3 Ge7" for the two individual detectors from the recent measurement. 
The M1 measurement was performed with a single HPGe detector called Ge4 for a live-time of \unit[170]{d}. The M2 measurement was performed for \unit[68]{d} with the same sample and Sandwich setup consisting of detectors Ge6 and Ge7 as used for the recent measurement but with their energy spectra combined as shown in Ref.~\cite{Ta108m2009}. The M3 measurement was performed for \unit[176]{d}.    

For the $\beta^-$ branch the single \gline\ at \unit[234.0]{keV} is used. The \unit[350.9]{keV} \gline\ is omitted because it cannot be clearly distinguished from the prominent \unit[351.9]{keV} background \gline\ from \nuc{Pb}{214} with the present energy resolution. The \unit[103.5]{keV} \gline\ has a small emission probability and small detection efficiency and is thus neglected. For the EC branch also the \unit[93.4]{keV} \gline\ is neglected due to the same argument. The fit for this branch is based on the \unit[215.4]{keV} and \unit[332.3]{keV} \glines. 
Each de-excitation \gline\ $k$ in a given decay mode has its own fit region of \unit[$\pm30$]{keV} around the \gline\ energy. Thus the $\beta^-$ branch has one fit region and the EC branch has two fit regions.\\

The signal counts $s_{d,k}$ of each \gline\ in each dataset are connected with the half-life $T_{1/2}$ of the decay mode as
\begin{eqnarray}
\label{eq:PdHLtoCounts}
s_{d,k} =
\ln{2} \cdot  \frac{1}{T_{1/2}} \cdot \epsilon_{d,k} \cdot N_A \cdot T_d \cdot m \cdot f  \cdot \frac{1}{M}\ ,
\end{eqnarray}

where $\epsilon_{d,k}$ is the full energy detection efficiency of \gline\ $k$ in dataset $d$, $
N_A$ is the Avogadro constant,
$T_d$ is the live-time of the dataset, 
$m$ is the mass of the Ta-sample,
$f$ is the natural isotopic abundance of \nuc{Ta}{180} and $M$ the molar mass of natural tantalum. 
%
The Bayesian Analysis Toolkit (BAT) \cite{Caldwell:2009kh} is used to perform a maximum posterior fit combining all four datasets and \glines\ for a given decay mode. The likelihood $\mathcal{L}$ is defined as the product of the Poisson probabilities of each bin $i$ in fit region $k$ in every dataset $d$ 
\begin{eqnarray}
\mathcal{L}(\mathbf{p}|\mathbf{n}) =
\prod \limits_d \prod \limits_k \prod \limits_i \frac{\lambda_{d,k,i}(\mathbf{p})^{n_{d,k,i}}}{n_{d,k,i}!} e^{-\lambda_{d,k,i}(\mathbf{p})}\ ,
\end{eqnarray}

where \textbf{n} denotes the data and \textbf{p} the set of floating parameters.
$n_{d,k,i}$ is the measured number of counts and $\lambda_{d,k,i}$ is the expected number of counts in bin $i$. $\lambda_{d,k,i}$ is taken as the integral of the extended p.d.f.\ $P_{d,r}$ in this bin
\begin{eqnarray}
\lambda_{d,k,i}(\mathbf{p}) &=&
 \int_{\Delta E_{d,k,i}} P_{d,k}(E|\mathbf{p}) dE\ , \label{eq:Ta_expcounts}
\end{eqnarray}

where $\Delta E_{d,k,i}$ is the bin width of \unit[0.5]{keV}. 
The counts in the fit region are expected from three different types of contributions which are used to construct $P_{d,k}$:
(1) a linear background, (2) the Gaussian signal peak and (3) a number of Gaussian background peaks. The number and type of background peaks depend on the fit region and will be described later. The full expression of $P_{d,k}$ is written as:
\begin{eqnarray}
P_{d,k}(E|\mathbf{p}) &&=
 B_{d,k} + C_{d,k}\left( E-E_0 \right) \label{eq:Ta_pdf}\\[2mm]
&&+  \frac{s_{d,k}}{\sqrt{2\pi}\sigma_{d,k}} 
\cdot \exp{\left(-\frac{(E-E_{k})^2}{2\sigma_{d,k}^2}\right)}\nonumber\\[2mm]
&&+ \sum \limits_{l_k} \left[\frac{b_{d,l_k}}{\sqrt{2\pi}\sigma_{d,k}} 
\cdot \exp{\left(-\frac{(E-E_{l_k})^2}{2\sigma_{d,k}^2}\right)}\right].\nonumber
\end{eqnarray}

The first line is describing the linear background with the two parameters $B_{d,k}$\ and $C_{d,k}$.
The second line is describing the signal peak with the energy resolution $\sigma_{d,k}$ and the \gline\ energy $E_k$.
The third line is describing the $l_k$ background peaks in fit region $k$ with the strength of the peak $b_{d,l_k}$ and the peak position $E_{l_k}$. The same p.d.f.\ with different parameter values is used for all four datasets.\\

The free parameters $\mathbf{p}$ in the fit for the $\beta^-$ (EC) branch are: 
\begin{itemize}
\item 1 (1) inverse half-life $(T_{1/2})^{-1}$ with flat prior
\item 8 (16) linear background parameters $B_{d,k}$ and $C_{d,k}$ with flat priors
\item 4 (8) energy resolutions $\sigma_{d,k}$ with Gaussian priors 
\item 4 (8) detection efficiencies $\epsilon_{d,k}$ with Gaussian priors 
\item 1 (2) signal peak positions $E_{k}$ with Gaussian priors
\item $l_1$ ($l_{1}$+$l_{2}$) background peak strength $b_{d,l_k}$ with flat priors 
\item $l_1$ ($l_{1}$+$l_{2}$) background peak positions $E_{l_r}$ with Gaussian priors
\end{itemize}

The energy resolutions are determined using  reference point sources including \nuc{Am}{241}, \nuc{Cs}{137} and \nuc{Co}{60}. The main gamma-lines of these radionuclides are fitted by a Gaussian distribution and the calculated energy resolutions are interpolated by a quadratic function.
The mean of the Gaussian priors is taken from these calibrations and reported in \tab \ref{tab:FWHM}. The width of these priors is taken as the uncertainty of the resolution calibration curve and approximated with \unit[5]{\%} for all datasets and \glines.\\

The full energy peak detection efficiencies as determined with the MC simulations are reported in \tab \ref{tab:efficiencies}. These values are taken as the mean value of the Gaussian prior.  
The uncertainty of the detection efficiencies is approximated with \unit[10]{\%} for each dataset and \gline\ and used as the width of the prior.\\

The posterior probability distribution is calculated from the likelihood and prior probabilities with BAT. The maximum of the posterior is the best fit. The posterior is marginalized for $(T_{1/2})^{-1}$ and used to extract the half-life limit with the \unit[90]{\%} quantile. This results in \unit[90]{\%} credibility limits on the half-life. Systematic uncertainties are included via the width of the Gaussian prior probabilities. 

\begin{table}
\caption{\label{tab:FWHM} Energy resolution in FWHM for all \glines\ and datasets. The uncertainty is taken as \unit[5]{\%} of the nominal value.}
\begin{ruledtabular}
\begin{tabular}{l|rrrr}
\gline\ energy  & M1 & M2 & M3\_Ge6 & M3\_Ge7   \\
\hline
$\beta^-$\\
					   \unit[234.0]{keV}  & \unit[1.56]{keV} & \unit[1.75]{keV} &  \unit[1.82]{keV} & \unit[1.50]{keV} \\
\hline
EC\\
					   \unit[215.3]{keV}  & \unit[1.54]{keV} & \unit[1.75]{keV} &  \unit[1.80]{keV} & \unit[1.49]{keV} \\
					   \unit[332.3]{keV}  & \unit[1.63]{keV} & \unit[1.85]{keV} &  \unit[1.91]{keV} & \unit[1.57]{keV} \\
\end{tabular}
\end{ruledtabular}
\end{table}

\begin{table}
\caption{\label{tab:efficiencies} Full energy peak detection efficiencies per decay of \nuc{Ta}{\rm 180m} into the $\beta^-$ or EC branch. The uncertainty on the efficiencies is taken as \unit[10]{\%} of the nominal value.}
\begin{ruledtabular}
\begin{tabular}{l|rrrr}
\gline\ energy  & M1 & M2 & M3\_Ge6 & M3\_Ge7   \\
\hline
$\beta^-$\\
					   \unit[234.0]{keV}  & \unit[0.54]{\%} & \unit[1.16]{\%} &  \unit[0.46]{\%} & \unit[0.66]{\%} \\
\hline
EC\\
					   \unit[215.3]{keV}  & \unit[0.45]{\%} & \unit[0.96]{\%} &  \unit[0.38]{\%} & \unit[0.55]{\%} \\
					   \unit[332.3]{keV}  & \unit[1.07]{\%} & \unit[2.40]{\%} &  \unit[0.96]{\%} & \unit[1.34]{\%} \\
\end{tabular}
\end{ruledtabular}
\end{table}

\section{Results}

\begin{figure*}
\includegraphics[width=0.99\textwidth]{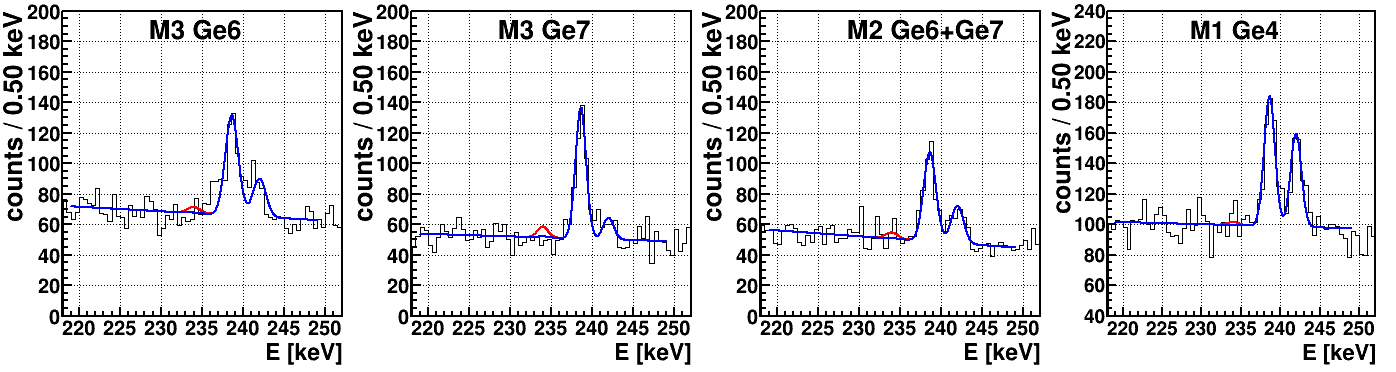}
\caption{\label{fig:pdf_ROI_Composition_Ta180m_bm} Region of interest for the $\beta^-$ channel of \nuc{Ta}{\rm 180m} decay in the four different datasets. The best fit is shown in blue and the best fit with the signal peak set to the \unit[90]{\%} C.I. half-life limit is shown in red.}
\end{figure*}

The single fit region for the $\beta^-$ branch is shown in \fig \ref{fig:pdf_ROI_Composition_Ta180m_bm}. Two background peaks are considered in the fit region at \unit[238.6]{keV} coming from \nuc{Pb}{212} (\unit[43.6]{\%}) and at \unit[241.0]{keV} coming from \nuc{Ra}{224} (\unit[4.1]{\%}) where the values in parenthesis denote the emission probability. The best fit is shown as blue line in the plots and finds zero counts for the signal process. The \unit[90]{\%} quantile of the marginalized posterior inverse half-life distribution yields a half-life limit of 
\begin{equation}
\beta^-:\hspace{1pc} T_{1/2} > 5.8 \cdot 10^{16}\, {\rm yr}\ (\unit[90]{\%}\ \rm C.I.)\ .
\end{equation}

This is illustrated in the plots as the red line where the signal peak strength is set to the half-life limit for each dataset.

The background rate underneath the signal peak is determined from the fit as \unit[1.53]{cts/keV/d} for Ge6 in M3, \unit[1.17]{cts/keV/d} for Ge7 in M3, \unit[2.99]{cts/keV/d} for M2 and \unit[2.35]{cts/keV/d} for M1. The comparison between the Sandwich setup in 2009 with \unit[2.99]{cts/keV/d} and in 2015 with \unit[2.70]{cts/keV/d} illustrates the background reduction over time by about 10\% e.g.\ due to the decay of \nuc{Ta}{182} (T$_{1/2}$ = 114.7~d). An investigation of the \nuc{Ta}{182} peaks at 1189.0~keV and 1221.4~keV found a $4.5\pm0.5$~mBq activity in M2 whereas only a $<0.23$~mBq (90\% CL) upper limit in M3. Also the background contributions from \nuc{Bi}{214} and \nuc{Pb}{214} were found to decrease from a weighted average of $3.7\pm1.2$~mBq in M2 to $2.2\pm1.6$~mBq in M3. See also Ref. \cite{Hult:2013} for more discussion on backgrounds for these detectors.\\

\begin{figure*}
\includegraphics[width=0.99\textwidth]{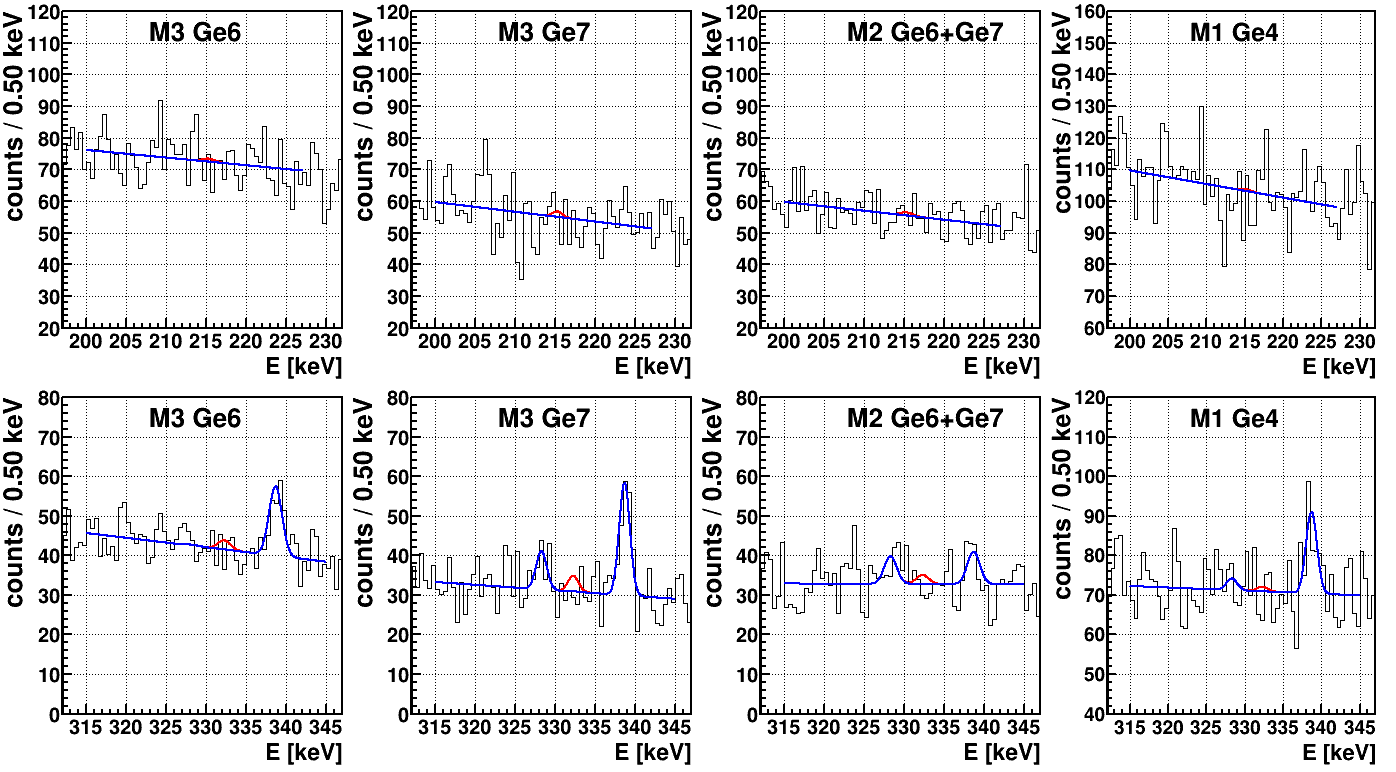}
\caption{\label{fig:pdf_ROI_Composition_Ta180m_ec} Regions of interest for the electron capture channel of \nuc{Ta}{\rm 180m} decay in the four datasets. The best fit is shown in blue and the best fit with the signal peak set to the \unit[90]{\%} C.I. half-life limit is shown in red.}
\end{figure*}

The two fit regions for the EC branch are shown in \fig \ref{fig:pdf_ROI_Composition_Ta180m_ec}. In the second fit region are two prominent background \glines\ at \unit[328.0]{keV} (\unit[3.0]{\%}) and \unit[338.32]{keV} (\unit[11.3]{\%}) from \nuc{Ac}{228} which are included in the fit. In the first fit region no background \gline\ is included. The best fit finds zero counts from the signal process which translates into a half-life limit of
\begin{equation}
{\rm EC}:\hspace{1pc} T_{1/2} > 2.0 \cdot 10^{17}\, {\rm yr}\ (\unit[90]{\%}\ \rm C.I.)\ .
\end{equation}

The background rate underneath the two signal peaks of \unit[215.3]{keV} and \unit[332.3]{keV} is \unit[1.65]{cts/keV/d} and \unit[0.95]{cts/keV/d} for Ge6 in M3, \unit[1.25]{cts/keV/d} and \unit[0.70]{cts/keV/d} for Ge7 in M3, \unit[3.28]{cts/keV/d} and \unit[1.92]{cts/keV/d} for M2 and \unit[2.42]{cts/keV/d} and \unit[1.67]{cts/keV/d} for M1, respectively. Also for this decay mode the background below the peaks decreased by about 10\%.\\

With these partial half-life limits, the total lower half-life limit of \nuc{Ta}{\rm 180m} is found as

\begin{equation}
{\rm ^{180m}Ta}:\hspace{1pc} T_{1/2} > 4.5 \cdot 10^{16}\, {\rm yr}\ (\unit[90]{\%}\ \rm C.I.)\ .
\end{equation}

\section{Conclusions}

A new investigation of the $\beta^-$ decay and EC branch of \nuc{Ta}{180\rm m} was performed with a Sandwich HPGe gamma spectroscopy setup in the HADES underground laboratory. A significant increase in exposure compared to previous measurements was achieved. In addition, the datasets of the old measurements were combined with the new datasets in a Bayesian framework incorporating various systematic uncertainties. The storage of the sample and the detector system underground for the last years decreased the background for this analysis by about 10\% compared to the previous measurement. No signal was observed for either decay mode and a new limit of \unit[$5.8\cdot10^{16}$]{yr} was set for the $\beta^-$ branch, \unit[$2.0\cdot10^{17}$]{yr} for the EC branch and \unit[$4.5\cdot10^{16}$]{yr} for the total half-life of \nuc{Ta}{\rm 180m}. Compared to the previously limits, this is an improvement of a factor of 1.6 and 5.5 for the $\beta^-$ and EC channel respectively and an improvement of a factor of 2.3 for the total half-life. The smaller improvement in the $\beta^-$ channel is mainly due to the fact that the \unit[350.9]{keV} cannot be effectively used in the spectral fit. \\
Additional improvements are difficult with this detector system but could be in principle achieved with increasing the target mass in an arrangement which does not decrease the detection efficiency. Further lowering the environmental background in a deeper underground location with more radiopure detector materials would also be beneficial.\\



\begin{acknowledgments}

This project was granted under the JRC-Geel's transnational access scheme, EUFRAT.  The HADES-staff of EURIDICE is gratefully acknowledged for their work.

\end{acknowledgments}

\appendix

\nocite{*}
\bibliography{aipsamp}

\end{document}